\documentclass[twocolumn,floatfix,superscriptaddress,a4paper,showpacs,showkeys,nofootinbib,reprint,prc]{revtex4-1}
\usepackage{epsfig}
\usepackage{latexsym}
\usepackage{xspace}
\usepackage[colorlinks=true,linktocpage=true,linkcolor=blue,citecolor=blue,allcolors=blue]{hyperref}
\usepackage[utf8]{inputenc}
\usepackage{indentfirst}
\usepackage{enumerate}
\usepackage{color}

\usepackage[caption=false,position=top]{subfig}

\usepackage{amsmath}
\usepackage{amssymb}
\usepackage[english]{babel}
\usepackage{url}
\newcommand{\eq}[1]{\begin{align} #1 \end{align}}

\begin{document}

%TC:ignore

\title{
Critical point signatures in the cluster expansion in fugacities
}

\author{Volodymyr Vovchenko}
\affiliation{
Institut f\"ur Theoretische Physik,
Goethe Universit\"at Frankfurt, Max-von-Laue-Str. 1, D-60438 Frankfurt am Main, Germany}
\affiliation{Frankfurt Institute for Advanced Studies, Giersch Science Center, Ruth-Moufang-Str. 1, D-60438 Frankfurt am Main, Germany}

\author{Carsten Greiner}
\affiliation{
Institut f\"ur Theoretische Physik,
Goethe Universit\"at Frankfurt, Max-von-Laue-Str. 1, D-60438 Frankfurt am Main, Germany}

\author{Volker Koch}
\affiliation{
Nuclear Science Division, Lawrence Berkeley National Laboratory, Berkeley, CA, 94720, USA}

\author{Horst Stoecker}
\affiliation{
Institut f\"ur Theoretische Physik,
Goethe Universit\"at Frankfurt, Max-von-Laue-Str. 1, D-60438 Frankfurt am Main, Germany}
\affiliation{Frankfurt Institute for Advanced Studies, Giersch Science Center, Ruth-Moufang-Str. 1, D-60438 Frankfurt am Main, Germany}
\affiliation{GSI Helmholtzzentrum f\"ur Schwerionenforschung GmbH, D-64291 Darmstadt, Germany}

\begin{abstract}
The QCD baryon number density can formally be expanded into a Laurent series in fugacity, which is a relativistic generalization of Mayer's cluster expansion.
We determine properties of the cluster expansion in a model with a phase transition and a critical point at finite baryon density, in which the Fourier coefficients of the expansion can be determined explicitly and to arbitrary order.
The asymptotic behavior of Fourier coefficients changes qualitatively as one traverses the critical temperature and it is connected to the branch points of a thermodynamic potential associated with the phase transition.
The results are discussed in the context of lattice QCD simulations at imaginary chemical potential.
We argue that the location of a branch point closest to the imaginary chemical potential axis can be extracted through an analysis of an exponential suppression of Fourier coefficients.
This is illustrated using the four leading coefficients both in a toy model as well as by using recent lattice QCD data.
\end{abstract}

%\pacs{24.10.Pa, 25.75.Gz}


\maketitle

%TC:endignore 

\section{Introduction}

Identification of the phases and structure of strongly interacting matter at finite baryon densities is one of the outstanding issues in modern nuclear physics. 
It has been established in the framework of lattice QCD that the quark-hadron transition at vanishing baryon density is a smooth crossover~\cite{Aoki:2006we}. 
On the other hand, it is expected~(although not proven) that a first-order quark-hadron transition takes place at sufficiently high baryon density, with the associated QCD critical point~(CP)~\cite{Stephanov:1998dy}.
The search for a critical behavior at finite $\mu_B$ is performed using measurements of fluctuations in heavy-ion collisions~\cite{Stephanov:2008qz,Koch:2008ia,Gazdzicki:2015ska,Luo:2017faz} and also using indirect lattice methods such as Taylor expansion around $\mu_B = 0$~\cite{Allton:2002zi,Gavai:2008zr,Kaczmarek:2011zz,Bazavov:2017dus} or analytic continuation from imaginary $\mu_B$~\cite{deForcrand:2002hgr,DElia:2002tig,Gunther:2016vcp}.
Currently, the available lattice QCD results show little (if any) hints for a CP~\cite{Bazavov:2017dus,Fodor:2018wul}.

In the present work we will consider the above questions in the framework of an expansion
\eq{\label{eq:clusterexpansion}
\frac{p}{T^4} = \frac{1}{VT^3} \ln Z = \frac{1}{2} \sum_{k=-\infty}^{\infty} \, p_{|k|} (T) \, e^{k \, \mu_B/T}~,
}
which represents the QCD grand canonical potential as a Laurent series in baryon number fugacity $\lambda_B \equiv e^{\mu_B/T}$.
Formally, it can be viewed as a relativistic extension of Mayer’s cluster expansion in fugacities~\cite{GNS}.\footnote{This expansion is often called the ``relativistic virial expansion'' in the literature~\cite{Venugopalan:1992hy}.}
The net baryon density, $\rho_B = (\partial p /\partial \mu_B)_T$, reads
\eq{\label{eq:rhoBcl}
\frac{\rho_B}{T^3} & = \frac{1}{2} \sum_{k=-\infty}^{\infty} \, b_{|k|} (T) \, e^{k \, \mu_B/T} \nonumber
\\
& = \sum_{k=1}^{\infty} \, b_k(T) \, \sinh \left( \frac{k \mu_B}{T} \right), \quad b_k \equiv k \, p_k.
}
The cluster expansion~\eqref{eq:rhoBcl} of net baryon density is particularly interesting in the context of lattice QCD simulations at imaginary $\mu_B$. Indeed, $\rho_B$ attains a form of a trigonometric Fourier series for a purely imaginary $\mu_B = i \, \theta_B \, T$~\cite{DElia:2002tig,Kratochvila:2006jx,Bornyakov:2016wld}. The cluster expansion coefficients $b_k$ become Fourier coefficients:
\eq{\label{eq:FourierDef}
b_k(T) = \frac{2}{\pi} \int_0^{\pi} \text{Im} \left[ \frac{\rho_B(T, i \theta_B \, T)}{T^3} \right] \, \sin(k \, \theta_B) \, d \theta_B~.
}
The four leading Fourier coefficients have been computed at the physical point in Ref.~\cite{Vovchenko:2017xad}.
Recent applications of the Fourier expansion method include a construction of a crossover equation-of-state for finite baryon densities~\cite{Vovchenko:2017gkg,Vovchenko:2018zgt,Vovchenko:2019vsm},  a determination of the net baryon number distribution in heavy-ion collisions at the LHC~\cite{Bzdak:2018zdg}, and analysis of the scaling properties of $b_k$ related to the chiral phase transition~\cite{Almasi:2018lok,Almasi:2019bvl} or repulsive interactions~\cite{Taradiy:2019taz}.

In the present work we explore how a a critical endpoint of a first-order phase transition at finite baryon density affects the properties of the cluster expansion, in particular the asymptotic behavior of the Fourier coefficients.
To this end, we develop a toy model containing a phase transition and a critical point for which one can evaluate \emph{all} coefficients $b_k$ of the cluster expansion explicitly.
The behavior of Fourier coefficients associated with the phase transition criticality is elaborated.
Based on universality argument, the results obtained are expected to be quite generic for any first-order phase transition with a critical endpoint at finite density. This is additionally demonstrated in Appendix B for a Nambu-Jona-Lasinio description.
We then explore a possibility of extracting the location of thermodynamic singularities from a number of leading Fourier coefficients and show that such a procedure is feasible under certain circumstances.

\section{Trivirial model}
\label{sec:tvm}

For simplicity, we consider first a single-component Maxwell-Boltzmann gas of interacting particles.
In the context of QCD, the particles can be regarded as abstract baryonic degrees of freedom. 
The scaled particle number density, $n / T^3$, has the following cluster expansion form
\eq{\label{eq:ncl}
\frac{n(T, \mu)}{T^3} = \frac{1}{2} \, \sum_{k=1}^{\infty} \, b_k(T) \, \lambda^k~.
}
The ideal gas limit corresponds to truncating the series at the first term. This is the case for the partial pressure of baryons in the ideal hadron resonance gas model.

The CP-symmetry of QCD can be recovered by adding an antisymmetric contribution of antibaryons to Eq.~\eqref{eq:ncl}. In such a case $b_k$ corresponds to the Fourier coefficients defined in Eq.~\eqref{eq:FourierDef}.
Here we would like to determine how a phase transition at finite density influences the behavior of $b_k$.
To achieve this goal, we are looking for a theory containing a phase transition where one can evaluate $b_k$ explicitly.

Before proceeding to a model calculation, it is worthwhile to point out the expected large-$k$ behavior of $b_k$ based on generic features of power series expansions.
The series~\eqref{eq:ncl} converges for all complex values of $\lambda$ inside a circle around the origin which has a radius of $|\lambda_r|$. $|\lambda_r|$ is the radius of convergence of the power series~\eqref{eq:ncl}, which corresponds to the distance from the origin to the nearest point $\lambda_r$ in the complex $\lambda$ plane where $n/T^3$ cannot be defined as a holomorphic function of $\lambda$.
We will refer here to such a point as a singularity.
This singularity is located on the above-mentioned circle of radius $|\lambda_r|$.
The radius of convergence is encoded in the asymptotic behavior of the expansion coefficients. The general definition of $|\lambda_r|$ is
\eq{\label{eq:convr}
%|\lambda_r| = \underset{k \to \infty}{\lim \inf} \, \left( \frac{1}{2} |b_k| \right)^{-1/k}~.
|\lambda_r| = \left[\underset{k \to \infty}{\lim \sup} \, \left( \frac{1}{2} |b_k| \right)^{1/k}\right]^{-1}~.
}
A simple possibility which satisfies~\eqref{eq:convr} is an exponential dependence of $b_k$ on $\lambda_r$ in the large-$k$ limit:
\eq{\label{eq:bkexp}
b_k \overset{k \to \infty}{\sim} \lambda_r^{-k}.
}
Below we demonstrate the validity of Eq.~\eqref{eq:bkexp} in an explicit model calculation.

\subsection{Model definition}

Perhaps the simplest theory with a critical point of a first-order phase transition is the van der Waals~(vdW) equation, which is given in terms of the pressure as a function of the temperature and particle number density:
\eq{\label{eq:Pvdw}
p^{\rm vdW}(T,n) = \frac{T \, n}{1 - b \, n} - a \, n^2.
}
The grand canonical formulation of the vdW equation was considered in Refs.~\cite{Vovchenko:2015xja,Vovchenko:2015uda,Bzdak:2018uhv} to study particle number fluctuations associated with criticality.
On the other hand, an explicit vdW model determination of the coefficients $b_k(T)$ to arbitrary order $k$ does not appear to be straightforward.
For this reason we consider here a slightly different model, which is obtained by expanding the vdW equation~\eqref{eq:Pvdw} in power series in $n$ and truncating the series at the 3rd order:
\eq{\label{eq:Pmod}
p(T,n) = T \, n + T \, \left(b - \frac{a}{T}\right) n^2 + T \, b^2 \, n^3~.
}

\begin{figure}[t]
  \centering
  \includegraphics[width=.48\textwidth]{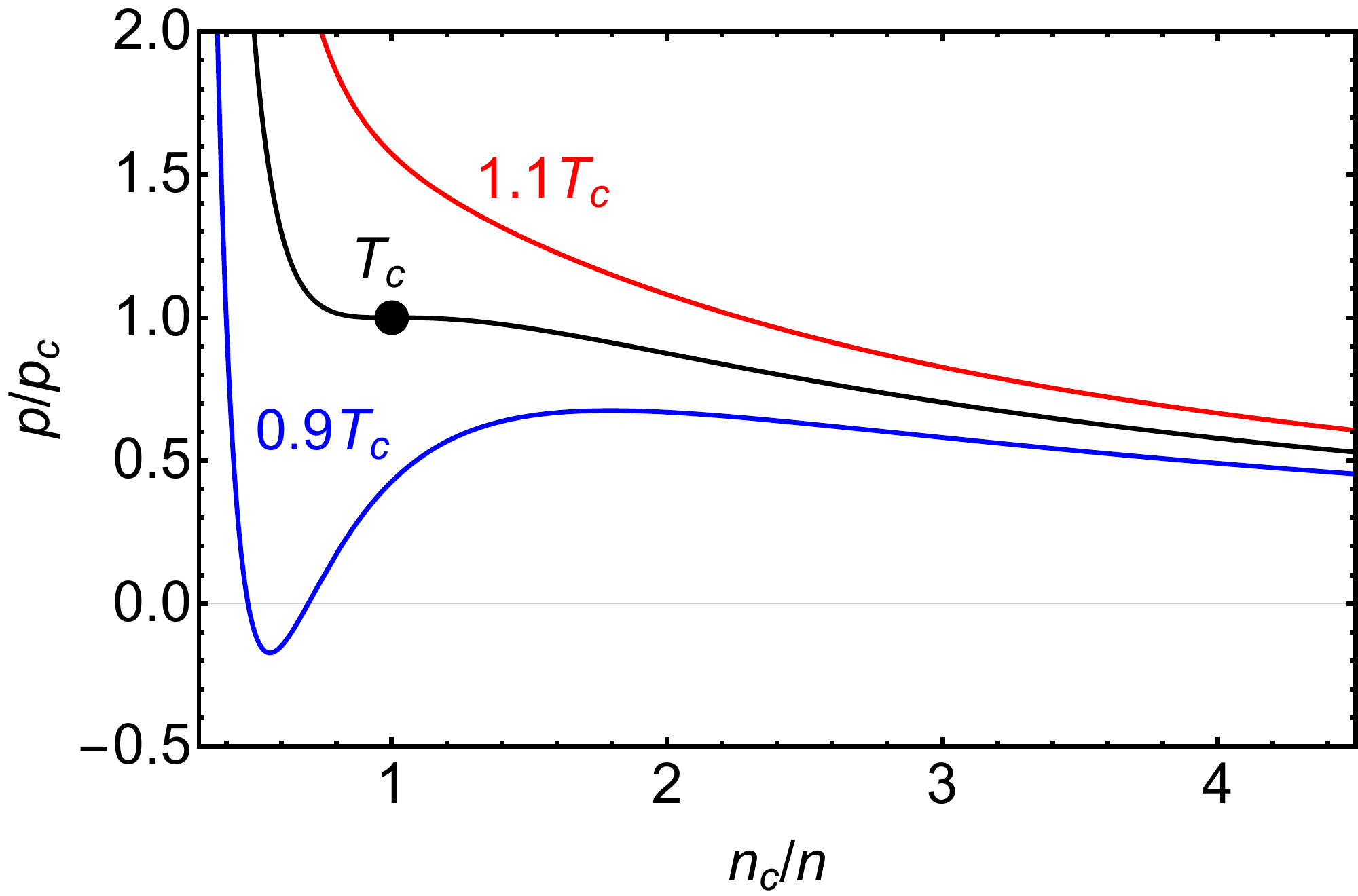}
  \caption{
   Pressure versus specific volume~(inverse density) isotherms for the trivirial model in reduced units. The symbol depicts the critical point.
  }
  \label{fig:isotherms}
\end{figure}

In this model the pressure is represented as a third-order polynomial in the particle number density. For this reason we will call this equation of state the \emph{trivirial model}~(TVM).
The qualitative behavior of the TVM isotherms coincides with the one in the standard vdW model: above a certain critical temperature $T_c$ the pressure isotherms are monotonically decreasing functions of the specific volume $v = n^{-1}$ while at $T < T_c$ they contain non-monotonic wiggles~(see Fig.~\ref{fig:isotherms}).
This implies an existence of a first-order liquid-gas transition with a critical point~(CP) in the TVM. The CP location is determined from  equations $(\partial p / \partial n)_T = 0$ and $(\partial^2 p / \partial n^2)_T = 0$:
\eq{\label{eq:CPloc}
T_c = \frac{\sqrt{3}-1}{2} \, \frac{a}{b}, \quad n_c = \frac{1}{\sqrt{3} \, b}, \quad p_c = \frac{3 - \sqrt{3}}{18} \, \frac{a}{b^2}~,
}
where we picked only the solution with $T_c > 0$.

\subsection{Grand canonical ensemble}

Equation~\eqref{eq:Pmod} defines the model pressure in the canonical ensemble. As $T$ and $n$ are not the natural variables of the pressure function, Eq.~\eqref{eq:Pmod} so far does not define the thermodynamic potential. This is achieved through a transformation to the grand canonical ensemble (GCE).
This will allow the analysis of the cluster expansion coefficients $b_k$.
In order to achieve that, we follow the procedure done in Ref.~\cite{Vovchenko:2015xja}. First, the free energy $F(T,V,N)$ is determined from the thermodynamic relation $p = - (\partial F / \partial V)_{T,N}$.
Integrating the pressure function~\eqref{eq:Pmod} and requiring that the free energy reduces to that of an ideal gas in the limit $N/V \to 0$ one obtains
\eq{\label{eq:F}
& F(T,V,N)  \nonumber \\
 & \quad = - T \, N  \left\{ 1 + \ln \frac{V \, \phi(T) \, e^{-b^2 N^2 / (2V^2) - b N / V} }{N}
 \right\} 
\nonumber \\
 & \qquad - \frac{a \, N^2}{V}~,
}
with
\eq{
\phi(T) & = \frac{d \, m^2 \, T}{2 \pi^2} \, K_2(m/T).
}
Here $d$ and $m$ are particle's degeneracy factor and mass, respectively, and $K_2$ is the
modified Bessel function.

The chemical potential, $\mu \equiv (\partial F / \partial N)_{T,V}$, reads
\eq{\label{eq:mu}
\mu = -T \, \ln[\phi(T)/n] + T \left[ \frac{3}{2} (bn)^2 + 2 \, b n \, \left(1 - \frac{a}{bT}\right) \right]~.
}

The fugacity reads
\eq{\label{eq:fug}
\lambda = \frac{n}{\phi(T)} \, \exp\left[ \frac{3}{2} (bn)^2 + 2 \, b n \, \left(1 - \frac{a}{bT}\right) \right]~.
}

Equation~\eqref{eq:fug}~[or, equivalently,~\eqref{eq:mu}] defines the particle number density~$n(T,\mu)$ in the GCE, i.e. density as a function of $T$ and $\mu$.
Substituting $n(T,\mu)$ into Eq.~\eqref{eq:Pmod} then allows one to reconstruct the GCE pressure.
Relation~\eqref{eq:fug} is a transcendental equation for the GCE density.

At given values of $T$ and (complex)~$\mu$, Eq.~\eqref{eq:fug} may have more than a single solution, meaning that $n(T,\mu)$ is a multivalued function. This multi-valuedness translates to the analytic properties of the GCE thermodynamic potential $p(T,\mu)$ and is expected to determine the convergence properties of the cluster expansion~\eqref{eq:ncl}.

\subsection{Branch points}

The particle number density $n$, as defined by Eq.~\eqref{eq:fug}, is a multi-valued function of the (complex) fugacity $\lambda$ at a fixed temperature $T$. Therefore, $n(\lambda;T)$ has branch points, which correspond to the zeroes of the derivative of the inverse function, i.e. $(\partial \lambda / \partial n)_T = 0$. 
Thus, $\lambda_{\rm br} = \lambda(n_{\rm br})$ are the branch points where $n_{\rm br}$ satisfies
\eq{\label{eq:betabr}
3 (b \, n_{\rm br})^2 + 2 \, \left(1 - \frac{a}{bT}\right) \, b \, n_{\rm br} + 1 = 0,
}
with the solution
\eq{\label{eq:betabr2}
n_{\rm br1,2} = \frac{\nu - 1 \pm \sqrt{(\nu - \nu_c)(\nu + \sqrt{3} - 1)} }{3b}.
}
Here $\nu \equiv a / (bT)$ and $\nu_c = a / (b T_c) = 2 / (\sqrt{3} - 1)$.

For the subcritical temperatures, $T<T_c$, the branch points $n_{\rm br1,2}$ are real, whereas for the supercritical temperatures, $T > T_c$, they correspond to a pair of complex conjugate numbers.
Expressing $\nu = \nu_c + \Delta \nu$ where $\Delta \nu < 0$ for $T > T_c$ and $\Delta \nu > 0$ for $T < T_c$, one has
\eq{\label{eq:betabr3}
n_{\rm br1,2} = \frac{\sqrt{3}+\Delta \nu}{3b} \pm \frac{\sqrt{(\sqrt{3} + \Delta \nu)^2 - 3}}{3b}.
}
A more detailed look shows that the two real roots at $T < T_c$ correspond to the spinodal points on the isotherms, see the blue line in Fig.~\ref{fig:isotherms}. 
In fact, one can see that Eq.~\eqref{eq:betabr} is equivalent to the equation $\partial \, p(T,n) / \partial n = 0$ defining the spinodal points.
These points correspond to the boundaries separating the mechanically unstable part of the isotherm~($\partial p /\partial n < 0$) from the phases of metastable gas~($n_{\rm br1}$) and metastable liquid~($n_{\rm br2} > n_{\rm br1}$).
At $T = T_c$ the two roots become degenerate, this corresponds to the critical point.
At $T > T_c$ the two complex conjugate roots are located in the complex plane away from the real axis. This corresponds to the so-called crossover singularities~\cite{Stephanov:2006dn}.
A similar behavior of thermodynamic singularities associated with a phase transition has earlier been reported for the vdW equation of state~\cite{hemmer1964yang}.

The locations of the branch points in the fugacity plane are the following
\eq{\label{eq:zbr}
\lambda_{\rm br1,2} = \frac{n_{\rm br1,2}}{\phi(T)} \, \exp\left[ -\frac{1}{2} - (\Delta \nu + \sqrt{3}) \, b \, n_{\rm br1,2} \right],
}
where we used Eq.~\eqref{eq:betabr}.
For the complex chemical potential plane one has:
\eq{\label{eq:mubr}
\mu_{\rm br1,2} = T \, \ln(\lambda_{\rm br1,2}) + i \, 2 \pi \, T \, k, \qquad k \in \mathbb{Z}.
}
Here the second term appears due to the periodicity of the chemical potential in the imaginary axis direction.

\subsection{Cluster expansion coefficients}

Let us define the principal branch of the GCE density $n(T, \lambda)$ by the condition that $n$ is real  for real values of $\lambda$ and that it reduces to the ideal gas density in the dilute limit, i.e. $n \to \phi(T) \, \lambda$ as $\lambda \to 0$.
This principal branch can then be expressed in a Taylor series of the form~\eqref{eq:ncl} around $\lambda = 0$.
Here we determine the coefficients $b_k(T)$ of the expansion.

The function $n(T,\lambda)$ is defined by Eq.~\eqref{eq:fug} implicitly, namely as the inverse of the function $\lambda(T,n)$.
Given that $\lambda(T,n)$ is analytic at the point $n = 0$ and $\lambda'_{n}(T,n=0) \neq 0$, one can apply the Lagrange inversion theorem~\cite{AbramowitzStegun} to evaluate the series coefficients $b_k$.
One obtains:
\eq{\label{eq:gk}
b_k & = 2 \frac{\phi(T)}{T} \, [b \, \phi(T)]^{k-1}
\frac{1}{k!} \nonumber \\
& \quad \times \lim_{\omega \to 0} \, \frac{d^{k-1}}{d \omega^{k-1}} \, \exp\left[-2 \, \left(1 - \frac{a}{bT}\right) \, k \, \omega  - \frac{3}{2} \, k \, \omega^2\right] .
}
Let us make a variable substitution $\omega = x \, \sqrt{2/(3k)}$:
\eq{\label{eq:gk2}
b_k & = 2 \frac{\phi(T)}{T} \, [b \, \phi(T)]^{k-1}
\frac{1}{k!} \, \left( \frac{3k}{2} \right)^{\frac{k-1}{2}} \nonumber \\
& \quad \times \lim_{\omega \to 0} \, \frac{d^{k-1}}{d x^{k-1}} \, \exp\left[-2 \, \sqrt{\frac{2k}{3}} \left(1 - \frac{a}{bT}\right) \, x - x^2\right] .
}

One can recognize the generating function of Hermite polynomials in the r.h.s. of Eq.~\eqref{eq:gk2},
\eq{
\exp \left( 2 \, t \, x - x^2 \right) = \sum_{n=0}^\infty H_n(t) \, \frac{x^n}{n!}~,
}
with $t = - [1 - a/(bT)] \, \sqrt{2k/3}$.
The higher-order derivatives evaluated at $x = 0$ in the r.h.s. of Eq.~\eqref{eq:gk2} correspond to the Taylor coefficients of the generating function of Hermite polynomials. One obtains
\eq{\label{eq:bksol}
b_k(T) & = 2 \, \frac{\phi(T)}{T^3} \, [b \, \phi(T)]^{k-1} \, \frac{1}{k!} \, \left( \frac{3k}{2} \right)^{\frac{k-1}{2}} \nonumber \\
& \quad \times
H_{k-1}\left[ -\sqrt{\frac{2k}{3}} \, \left(1-\frac{a}{b\,T} \right) \right]~.
}
The  four leading coefficients read
\eq{
b_1(T) & = 2 \, \frac{\phi(T)}{T^3}, \\
b_2(T) & = -4 \, \frac{\phi(T)}{T^3} \, b \phi(T) \, \left(1 - \frac{a}{bT}\right)~, \\
b_3(T) & = 9 \, \frac{\phi(T)}{T^3} \, [b \phi(T)]^2 \, \left[1 - \frac{8}{3} \, \frac{a}{bT} + \frac{4}{3} \, \left( \frac{a}{bT}\right)^2 \right]~, \\
\label{eq:bfour}
b_4(T) & = -\frac{56}{3} \,  \frac{\phi(T)}{T^3} \, [b \phi(T)]^3 \, \left(1 - \frac{a}{bT}\right) \nonumber \\
& \quad \times \left[1 - \frac{32}{7} \, \frac{a}{bT} + \frac{16}{7} \, \left( \frac{a}{bT}\right)^2 \right]~.
}

The first three coefficients evaluated in the TVM coincide with the vdW model result~(see Appendix in Ref.~\cite{Vovchenko:2017xad}). This is not surprising as the pressures in the two models coincide up to the 3rd power of the particle number density. Starting from the 4th coefficient~\eqref{eq:bfour}, however, the two models differ.

\subsection{Asymptotic behavior of $b_k$}

The asymptotic behavior of the cluster integrals $b_k$ determines the convergence properties of the cluster expansion~\eqref{eq:ncl}.
This behavior is determined by the asymptotic properties of Hermite polynomials, which are known.
Extra care should be taken here, as both the index and the argument of the Hermite polynomials in~\eqref{eq:bksol} tend to large values as $k \to \infty$.
In such a case the asymptotic behavior depends on the relative increase rate of the Hermite polynomial index and its argument.
These different behaviors were studied in Ref.~\cite{math/0601078}.
There are three cases relevant for our analysis:

\ \\
\begin{widetext}
\begin{enumerate}
    \item For $x > \sqrt{2n}$ the Hermite polynomials $H_n(x)$ admit the asymptotic representation~(Theorem 1 in~\cite{math/0601078})
    $$
    H_n(x) \stackrel{n \to \infty}{\simeq} \exp\left[ \frac{x^2 - \sigma x - n}{2} + n \, \ln(\sigma+x) \right] \, \sqrt{\frac{1}{2}\left(1+\frac{x}{\sigma}\right)}, \qquad \sigma = \sqrt{x^2 - 2n}.
    $$
    In our case $n = k-1$ and $x = -\sqrt{\frac{2k}{3}} \, \left(1 - \frac{a}{bT}\right)$.
    The condition $x > \sqrt{2n}$ corresponds to the subcritical temperatures, $T < T_c$.
    Recalling $\nu \equiv a / (bT)$ and setting $\nu = \nu_c + \Delta \nu$ one obtains  the following for the coefficients $b_k$:
    \eq{
    b_k & \sim k^{-3/2} \, \left\{  \frac{\sqrt{3}+\Delta \nu}{3} - \frac{\sqrt{(\sqrt{3} + \Delta \nu)^2 - 3}}{3} \right\}^{-k} \, \exp\left\{ k \, \left[ \frac{1}{2} + (\sqrt{3} + \Delta \nu) \frac{\sqrt{3}+\Delta \nu - \sqrt{(\sqrt{3} + \Delta \nu)^2 - 3}}{3}\right] \right\}~, \nonumber \\
    & \sim k^{-3/2} \, \left\{ b \, n_{\rm br1} \, \exp\left[ -\frac{1}{2} -  (\sqrt{3} + \Delta \nu) \, b \, n_{\rm br1} \right] \right\}^{-k}~, \nonumber \\
    & \sim \frac{(\lambda_{\rm br1})^{-k}}{k^{3/2}}~.
    }
    The asymptotic behavior of the cluster expansion coefficients has the form of an exponential damping superimposed on a power-law suppression.
    The exponential suppression is determined by the branch point $\lambda_{\rm br1}$, located on the real fugacity axis.~\footnote{It is noted in Ref.~\cite{Stephanov:2006dn} that the branch points do lie on the real axis at subcritical temperatures in the mean-field universality class, but that this fact may not necessarily extend to other universality classes.
    }

    \item For $x \approx \sqrt{2n}$ the Hermite polynomials $H_n(x)$ admit the asymptotic representation~(Theorem 3 in~\cite{math/0601078})
    $$
    H_n(x) \stackrel{n \to \infty}{\simeq} \exp\left[\frac{n}{2} \, \ln(2n) - \frac{3}{2} \, n + \sqrt{2n} \, x \right] \, \sqrt{2\pi} \, n^{1/6} \, \operatorname{Ai}\left[ \sqrt{2} \, (x - \sqrt{2n}) \, n^{1/6} \right].
    $$
    The notation $x \approx \sqrt{2n}$ here means that $x \to \sqrt{2n}$ in the limit $n \to \infty$, however $x$ can be different from $\sqrt{2n}$ for a finite value of $n$.
    The condition $x \approx \sqrt{2n}$ corresponds in the TVM
    %$\nu = \nu_c = 1 + \sqrt{3}$, i.e. 
    to the critical temperature, $T = T_c$. One obtains:
    \eq{\label{eq:gkcrit}
    b_k & \simeq \frac{b \, T_c^3}{2} \, [b\,\phi(T_c)]^{-k} \frac{3^{-7/6}}{\Gamma(2/3)} \, (\sqrt{3} \, e^{3/2})^k \, k^{-4/3} \simeq \frac{b \, T_c^3 \, 3^{-7/6}}{2 \, \Gamma(2/3)} \frac{(\lambda_c)^{-k}}{k^{4/3}}~.
    }
    The asymptotic behavior of $b_k$ at $T = T_c$ has the form of an exponential suppression superimposed on a power-law damping. The exponential part is determined by the critical fugacity value $\lambda_c$ which corresponds to the critical point location. 
    
    \item For $|x| < \sqrt{2n}$~(i.e. $T > T_c$), the Hermite polynomials $H_n(x)$ have the asymptotic representation~(Theorem 5 in~\cite{math/0601078})
    $$
    H_n\left[\sqrt{2n} \sin\theta\right] \stackrel{n \to \infty}{\simeq} \sqrt{\frac{2}{\cos\theta}} \, \exp \left\{ \frac{n}{2} \, \left[ \ln(2n) - \cos(2\theta) \right] \right\} \cos \left\{ n \left[ \frac{1}{2} \, \sin(2\theta) + \theta - \frac{\pi}{2} \right] + \frac{\theta}{2} \right\},
    $$
    where $-\frac{\pi}{2} < \theta < \frac{\pi}{2}$.
    Here $n = k-1$ and $\sin\theta = \sqrt{\frac{2k}{2k-2}} \, \left(1 + \frac{\Delta \nu}{\sqrt{3}}\right)$ with $\nu = \nu_c + \Delta \nu$, and $-\nu_c < \Delta \nu < 0$.
    First, one observes that the fugacity values $\lambda_{\rm br1,2}$~\eqref{eq:zbr} at the branch points of the thermodynamic potential can be written
    \eq{
    \lambda_{\rm br1,2} & = \frac{\exp\left[-\left(\frac{1}{2}+(\sin \theta_0)^2\right)\right]}{\sqrt{3} \, b \, \phi(T)} \, \exp\left[ \mp i \, \left(\theta_0 - \frac{\pi}{2} + \frac{\sin 2\theta_0}{2}\right) \right] \\
    & = |\lambda_{\rm br}| \, e^{\pm i \, \theta_{\rm br}}.
    }
    with
    $$
    \theta_0 = \arcsin{\left(1 + \frac{\Delta \nu}{\sqrt{3}}\right)}, \qquad \Delta \nu < 0~.
    $$
    One obtains for $b_k$:
    \eq{
    b_k & \sim \frac{|\lambda_{\rm br}|^{-k}}{k^{3/2}} \, \sin\left(k \, \theta_{\rm br} + \frac{\theta_0}{2}\right).
    }
    The asymptotic behavior of $b_k$ at $T>T_c$ corresponds to a damped oscillator superimposed on a power-law decay. 
    The branch points that define this behavior correspond to the so-called crossover singularities.
    Denoting $\lambda_{\rm br} = e^{\mu_{\rm br}/T}$ and $\mu_{\rm br} = \mu_{\rm br}^{\rm R} \pm i \, \mu_{\rm br}^{\rm I}$ one can see that $|\lambda_{\rm br}| = e^{\mu_{\rm br}^{\rm R}/T}$ and $\theta_{\rm br} = \mu_{\rm br}^{\rm I} / T$.
    Thus, the real part $\mu_{\rm br}^{\rm R}/T$ of the chemical potential at the branch point determines the exponential suppression of the magnitude of $b_k$ at $T>T_c$ whereas the imaginary part $\mu_{\rm br}^{\rm I}/T$ defines the period of oscillations.

\end{enumerate}
\end{widetext}

The obtained TVM results can be summarized as follows:
\eq{\label{eq:bkTmTc}
b_k & \simeq A_- \, \frac{e^{-\frac{k \, \mu_{\rm sp1}}{T}}}{k^{3/2}}, \qquad & T < T_c, \\
\label{eq:bkTTc}
b_k & \simeq A_c \, \frac{e^{-\frac{k \, \mu_{\rm c}}{T}}}{k^{4/3}}, \qquad & T = T_c, \\
\label{eq:bkTpTc}
b_k & \simeq A_+ \, \frac{e^{-\frac{k \, \mu_{\rm crs}^{\rm R}}{T}}}{k^{3/2}} \, \sin\left(k \frac{\mu_{\rm crs}^{\rm I}}{T} + \frac{\theta_0}{2} \right), \qquad & T > T_c.
}
Here $\mu_{\rm sp1}$ corresponds to the spinodal point of the first-order phase transition which delineates the metastable gaseous phase and the mechanically unstable phase at $T < T_c$, $\mu_c$ corresponds to the critical point at $T = T_c$, and $\mu_{\rm crs}^{\rm R}$ and $\mu_{\rm crs}^{\rm I}$ are, respectively, the real and imaginary parts of the chemical potential corresponding to the crossover branch points at $T > T_c$.

\begin{figure}[t]
  \centering
  \includegraphics[width=.48\textwidth]{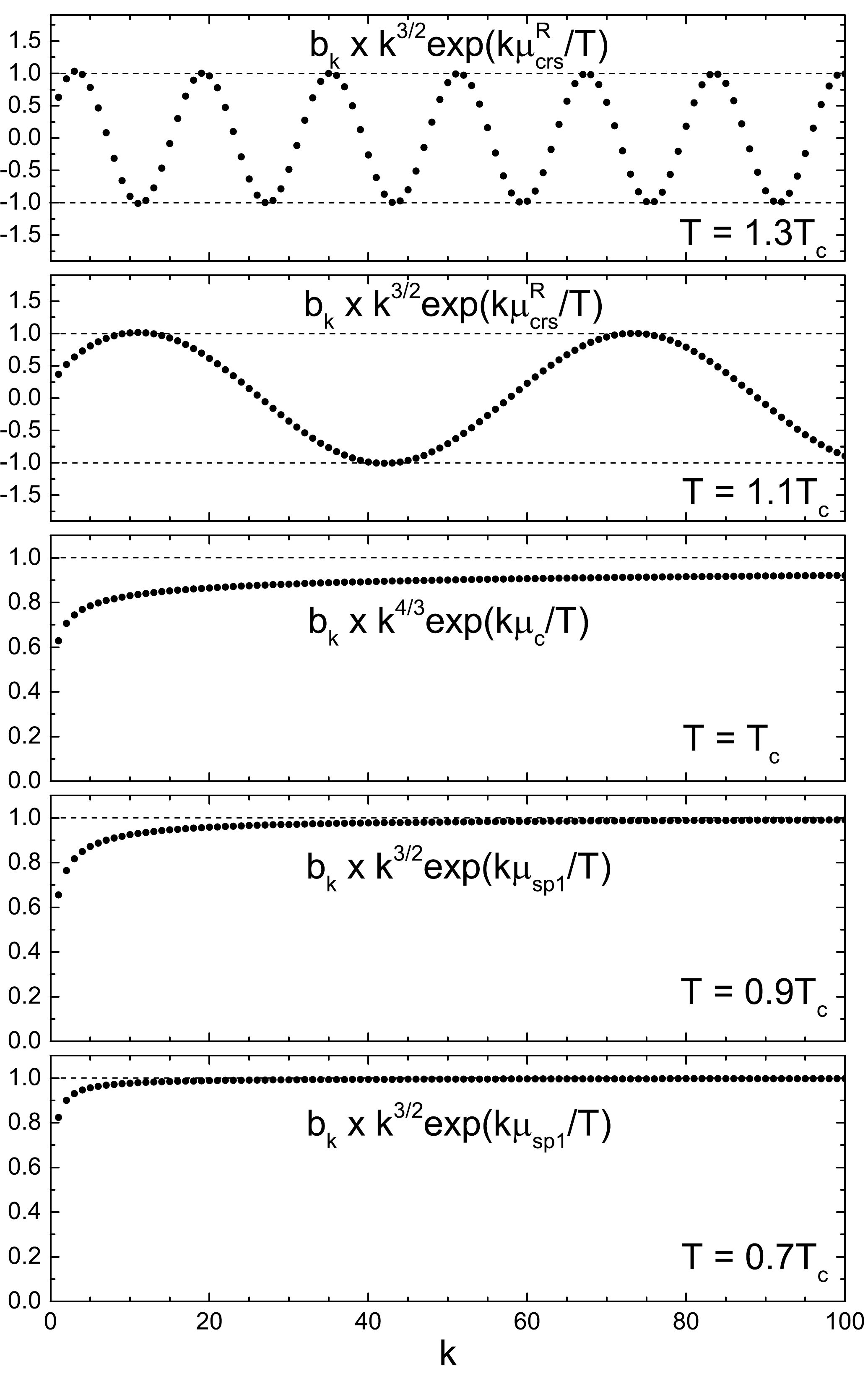}
  \caption{
   The $k$-dependence of the cluster expansion coefficients $b_k$ evaluated in the trivirial model using Eq.~\eqref{eq:bksol} for five different temperatures: $T = 1.3 \, T_c$, $T = 1.1 \, T_c$, $T = T_c$, $T = 0.9 \, T_c$, and $T = 0.7 \, T_c$~(from top to bottom).
   The coefficients are scaled by the expected asymptotic power-law, exponential, and amplitude factors~[Eqs.~\eqref{eq:bkTmTc}-\eqref{eq:bkTpTc}].
  }
  \label{fig:tvmbks}
\end{figure}

Figure~\ref{fig:tvmbks} depicts the $k$-dependence of $b_k$ in the TVM for five different temperatures: two temperatures above the critical one, $T = 1.3 \, T_c$ and $T = \, 1.1 T_c$, the critical temperature, $T = T_c$, and two temperatures below the critical one, $T = 0.7 \, T_c$ and $T = 0.9 \, T_c$.
The coefficients here are divided by the expected power-law, exponential, and amplitude factors from Eqs.~\eqref{eq:bkTmTc}-\eqref{eq:bkTpTc}.
We also set $b \phi(T) = 1$ in this calculation.
The large $k$ behavior of the computed coefficients is consistent with the expected asymptotics.
For $T < T_c$ the asymptotic behavior is approached monotonically. 
The rate of approach depends on the distance of an isotherm to the critical one.
For example, the reduced coefficients are within 10\% of the asymptotic limit at $k = 7$ for $T = 0.9 T_c$ and already at $k = 2$ for $T = 0.7T_c$.
%and fairly quickly: the reduced coefficients are within 10-15\% of the asymptotic limit already at $k = 5$.
For $T = T_c$ the large-$k$ limit is reached considerably slower.
For $T > T_c$ the coefficients exhibit an oscillatory behavior.
The period of oscillations is large at temperatures slightly above the critical one and decreases with increasing temperatures.
This behavior reflects the increase of the imaginary part $\mu_{\rm br}^{\rm I}/T$ of the crossover singularity chemical potential with the temperature at $T > T_c$, in accordance with Eq.~\eqref{eq:bkTpTc}.

The asymptotic behavior~\eqref{eq:bkTmTc}-\eqref{eq:bkTpTc} has been obtained here in the framework of the TVM.
Nevertheless, due to a universality of the critical behavior this result is expected to be the same for any theory with a phase transition and a critical point at finite density which belongs to the mean-field universality class, with nonuniversal constants $A_-$, $A_c$, $A_+$, and $\theta_0$.
In particular, we checked numerically that this holds for the vdW model~[Eq.~\eqref{eq:Pvdw}], through an evaluation of a large number of leading $b_k$ coefficients in that model.
Additionally, in Appendix~B we analyze the behavior of $b_k$  in a Nambu-Jona-Lasinio~(NJL) model through numerical calculations at an imaginary chemical potential. These calculations confirm that the asymptotic behavior~\eqref{eq:bkTmTc}-\eqref{eq:bkTpTc} also holds in NJL.

It should be noted that a phase diagram of a statistical system may contain richer structures than those given solely by a critical point that are studied here.
These could be, for example, a tricritical point and a line of second-order phase transitions, as expected for QCD in the chiral limit~\cite{Pisarski:1983ms,Halasz:1998qr}, or inhomogeneous phases~\cite{Deryagin:1992rw,Shuster:1999tn,Nakano:2004cd}.
How the phase structures of such systems are related to the $b_k$ asymptotics  is not obvious.
The TVM presented here is not suitable in such cases, as the model is only suited to determine features associated with a critical endpoint of a first-order phase transition.
Therefore, an analysis of Fourier coefficients within other manageable models possessing these involved phase structures is an interesting future possibility. One possible choice is the Gross-Neveu model in large $N_f$ limit~\cite{Gross:1974jv}, which has been used in the past to test proposals to analyze the phase structure of QCD~\cite{Hands:1992ck}, in particular using imaginary chemical potentials~\cite{Karbstein:2006er}.

A behavior of Fourier coefficients of net baryon density similar to Eqs.~\eqref{eq:bkTmTc}-\eqref{eq:bkTpTc} have recently been obtained in Ref.~\cite{Almasi:2019bvl} in the framework of Landau theory of phase transitions applied to the chiral phase transition in the limit of small quark masses, as well as using scaling relations.
One notable difference to the present work is a pre-exponential factor of $k^{-2}$ in the case of a chiral crossover in the mean-field approximation, which is different from the $k^{-3/2}$ factor obtained here~[see Eq.~\eqref{eq:bkTpTc}].
The apparent reason for this difference is that Ref.~\cite{Almasi:2019bvl} considers the chiral criticality at $\mu_B = 0$, where the perturbation in $\mu_B$ is coupled to the temporal Ising variable $t$, but where a coupling of $\mu_B$ to the magnetic field variable $h$ is forbidden by symmetry.
For a critical point at finite baryon density the situation is different: $\mu_B$ is coupled to both the $t$ and $h$ Ising variables, and the variable $h$ is expected to dominate the scaling near the critical point~\cite{Stephanov:2006dn}.
This behavior is reflected in the TVM, leading to the pre-exponential factor $k^{-3/2}$ instead of $k^{-2}$.

Another important remark is related to the universality class of the critical behavior associated with a critical point of the phase transition. 
As mentioned above, in the TVM this is the mean-field universality class.
The expected universality class for the QCD critical point is $Z(2)$~(3D-Ising)~\cite{Berges:1998rc,Halasz:1998qr}, which is characterized by somewhat different critical exponents~\cite{GNS}.
Therefore, for a universality class different from mean-field one expects a similar asymptotic behavior to the one given in Eqs.~\eqref{eq:bkTmTc}-\eqref{eq:bkTpTc}, but with corrections to the power-law exponents.

Based on the considerations above, we expect the following asymptotic behavior of $b_k$ in a general case:
\eq{\label{eq:bkgen}
b_k(T) \simeq A \, \frac{e^{-\frac{k \, \mu_{\rm br}^{\rm R}}{T}}}{k^{\alpha}} \, \sin\left(k \frac{\mu_{\rm br}^{\rm I}}{T} + \theta \right).
}
Here $\mu_{\rm br} = \mu_{\rm br}^{\rm R} \pm i \, \mu_{\rm br}^{\rm I}$ is a singularity~(a branch point)
of the thermodynamic potential which determines the asymptotic behavior of the cluster expansion coefficients. 
Evidently, this has to be the singularity located the closest to the imaginary $\mu_B$ axis, as contributions from all other singularities will have a stronger exponential suppression, rendering their contributions to $b_k$ subleading.
This singularity may not necessarily be connected to a critical point of a phase transition at finite density studied here.

The exponent $\alpha$ depends on the nature of the singularity~(universality class, critical point, spinodal or crossover, etc.).
While the only singularities in the TVM are those related to the phase transition, in a more general case the form~\eqref{eq:bkgen} can also accommodate singularities not related to physical phase transitions~(see Ref.~\cite{Taradiy:2019taz} for a number of examples).

\section{Extracting thermodynamic singularities from Fourier coefficients}
\label{sec:fitting}

The TVM introduced above can be used to model a hypothetical phase transition and a critical point at finite baryon density.
In such a case one can associate interacting particles in the TVM with abstract baryonic degrees of freedom.
The net baryon density reads\footnote{Here we consider the simplest relativistic generalization of the TVM, where baryon-antibaryon interaction terms are neglected.}
\eq{
\rho_B = n_B - n_{\bar{B}},
}
where
\eq{
\frac{n_{B(\bar{B})}}{T^3} = \frac{1}{2} \sum_{k=1}^{\infty} b_k(T) \, \lambda_B^{\pm 1},
}
which implies
\eq{\label{eq:rhoBsym}
\frac{\rho_{B(\bar{B})}(T,\lambda_B)}{T^3} = \sum_{k=1}^{\infty} \, b_k(T) \, \sinh \left( \frac{k \mu_B}{T} \right).
}
The form~\eqref{eq:rhoBsym} coincides with the relativistic cluster expansion~[Eq.~\eqref{eq:rhoBcl}] meaning that $b_k$ correspond to the Fourier coefficients of net baryon density at imaginary $\mu_B$.
The large $k$ behavior of Fourier coefficients associated with a phase transition at finite baryon density is given by Eqs.~\eqref{eq:bkTmTc}-\eqref{eq:bkTpTc}. 

Leading Fourier coefficients can in principle be calculated using lattice QCD simulations at imaginary $\mu_B$ through the Fourier transform.
In fact, this has already been done for the four leading coefficients in Ref.~\cite{Vovchenko:2017xad} on $N_{\tau} = 12$ lattices.
The question that we want to address is the following: can one extract useful information about QCD thermodynamic singularities from a number of leading Fourier coefficients based on the known expected asymptotic behavior?
We argue that the answer to this question is affirmative.
Moreover, we show that some useful information can be extracted from the already available lattice data.

Based on the general asymptotics~\eqref{eq:bkgen} of Fourier coefficients one notices that by far the strongest effect on the overall magnitude of Fourier coefficients is exerted by the real part $\mu_{\rm br}^{\rm R}$ of the branch point closest to the real $\mu_B$ axis.
More specifically, one has
\eq{\label{eq:bkasym}
\ln |b_k| \lesssim \ln A - \alpha \, \ln k - \frac{\mu_{\rm br}^R}{T} \, k~,
}
where the strongest $k$-dependence is in the third term.
It can be reasonable to expect the appearance of strong exponential suppression of Fourier coefficients already in the leading coefficients.
If that is the case, $\mu_{\rm br}^R$ can be extracted by fitting the absolute magnitudes of a number of the leading $b_k$'s with an ansatz
\eq{\label{eq:fitansatz}
\ln |b_k| = \ln A - \alpha \, \ln k - \frac{\mu_{\rm br}^R}{T} \, k~.
}

\begin{figure}[t]
  \centering
  \includegraphics[width=.48\textwidth]{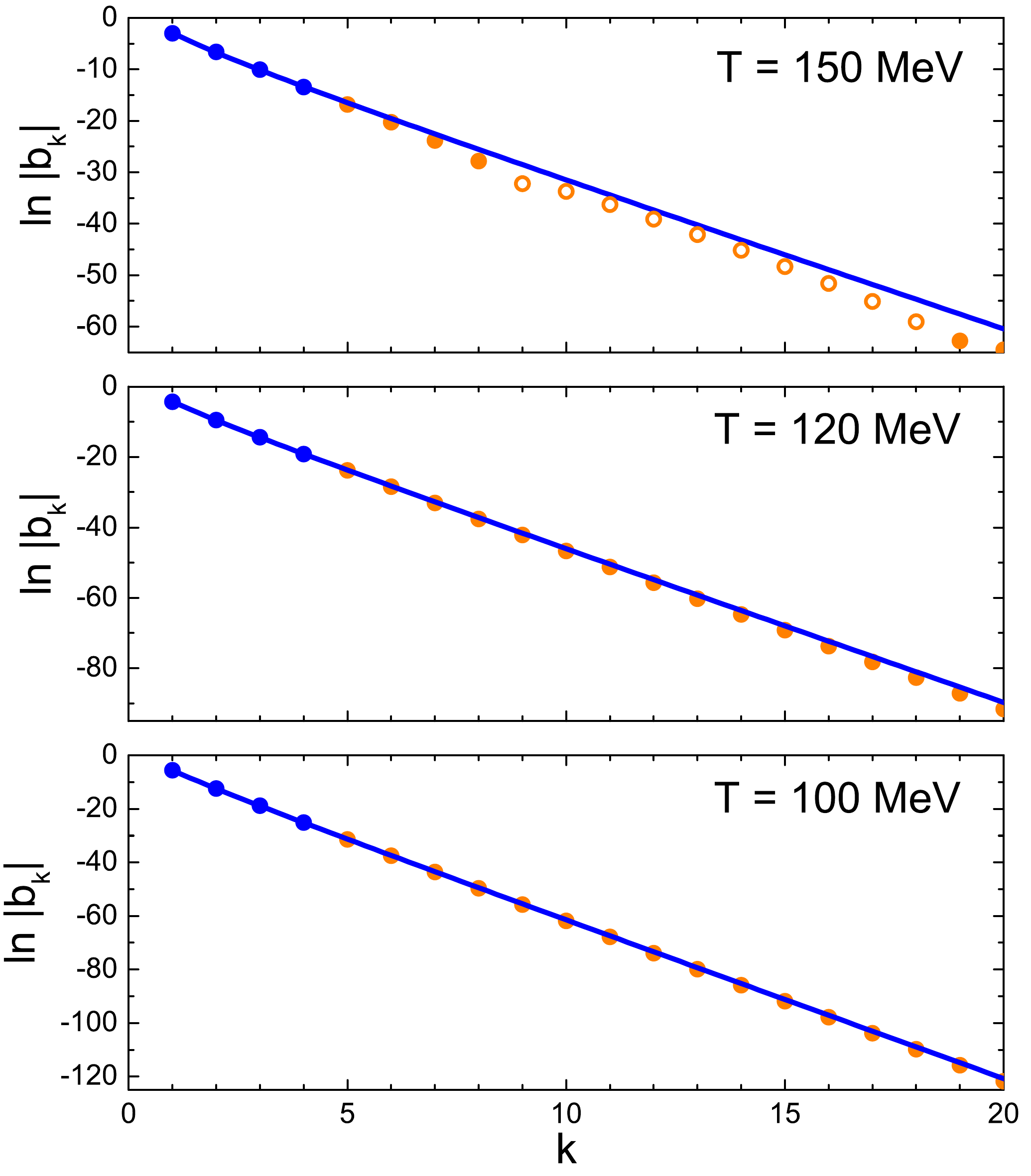}
  \caption{
   The blue lines depict the results of the fits to the four leading Fourier coefficients of the TVM with the ansatz~\eqref{eq:fitansatz} for three different temperatures: $T = 100$~MeV~(lower panel),
   $T \simeq T_c = 120$~MeV~(middle panel),
   and $T \simeq T = 150$~MeV~(upper panel).
   The blue points correspond to the four leading Fourier coefficients that were used in the fitting procedure, while the orange points depict the higher-order Fourier coefficients that were not used in the fits.
   The full and open symbols corresponds to positive and negative values of $b_k$, respectively.
  }
  \label{fig:tvmfits}
\end{figure}

As a proof of concept, we take the TVM for baryons with model parameters fixed in such a way as to obtain a critical point at $T_c = 120$~MeV and $\mu_c = 528$~MeV~($a = 328$~MeV fm$^3$, $b = 1$~fm$^3$, $d = 10$, and $m = 938$~MeV/$c^2$).
We fit the four leading Fourier coefficients in the TVM with the ansatz~\eqref{eq:fitansatz} at three temperatures: $T = 100$~MeV, $T \approx T_c = 120$~MeV, and $T = 150$~MeV. 
Parameters $A$ and $\mu_{\rm br}^R$ are fitted while $\alpha$ is fixed to its expected value of $3/2$.

The fit results are depicted in Fig.~\ref{fig:tvmfits} by the blue lines for the $k$-dependence of $\ln |b_k|$.
The fitted function provides a reasonable description of the higher-order $b_k$'s that were not used in the fitting procedure.
This is especially the case for $T = 100$~MeV and $T = 120$~MeV.
The extracted values of $\mu_{\rm br}^R$ can be compared with exact values~[Eq.~\eqref{eq:mubr}]: 5.83 vs 5.89 at $T = 100$~MeV, 4.26 vs 4.40 at $T = 120$~MeV, and 2.79 vs 2.98 at $T = 150$~MeV.
A fit at $T \approx T_c = 120$~MeV with the power-law exponent $\alpha$ equal to $4/3$~(the expected value for the critical isotherm) instead of $3/2$, yields $\mu_{\rm br}^R = 4.34$, i.e. the procedure is not very sensitive to moderate variations in $\alpha$.
The extracted $\mu_{\rm br}^R$ values reproduce the true ones to a fairly good precision~(10\% or better), even at $T > T_c$ where the exponential suppression of $b_k$ is superimposed on an oscillatory behavior.
The procedure can be improved by including higher-order coefficients into the fit. Omitting a number of leading coefficients from the fit could be helpful as well, as the asymptotic form~\eqref{eq:bkasym} is less justified for these coefficients than for the higher-order ones.

\begin{figure*}[t]
  \centering
  \includegraphics[width=.36\textwidth]{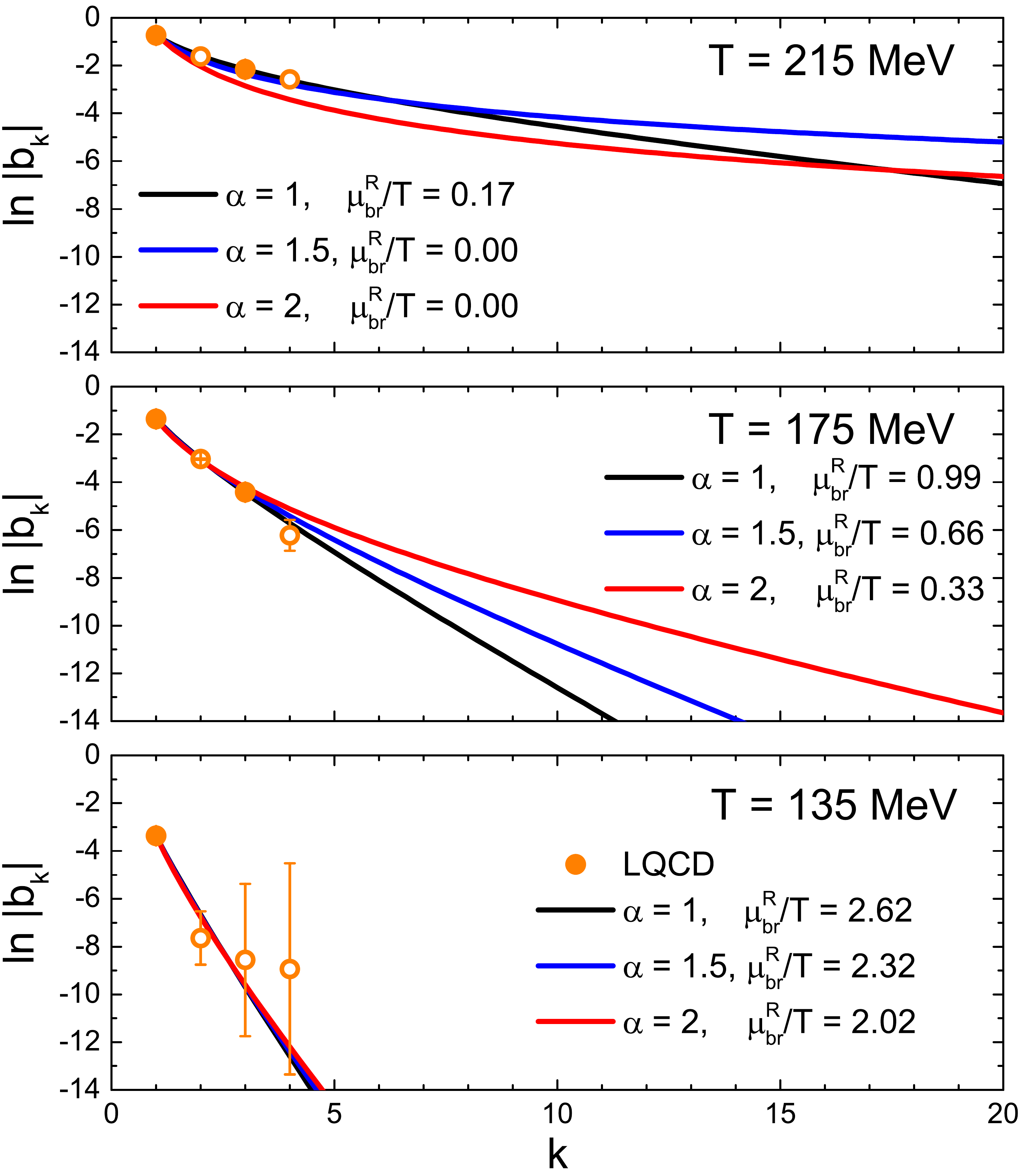}
  \includegraphics[width=.62\textwidth]{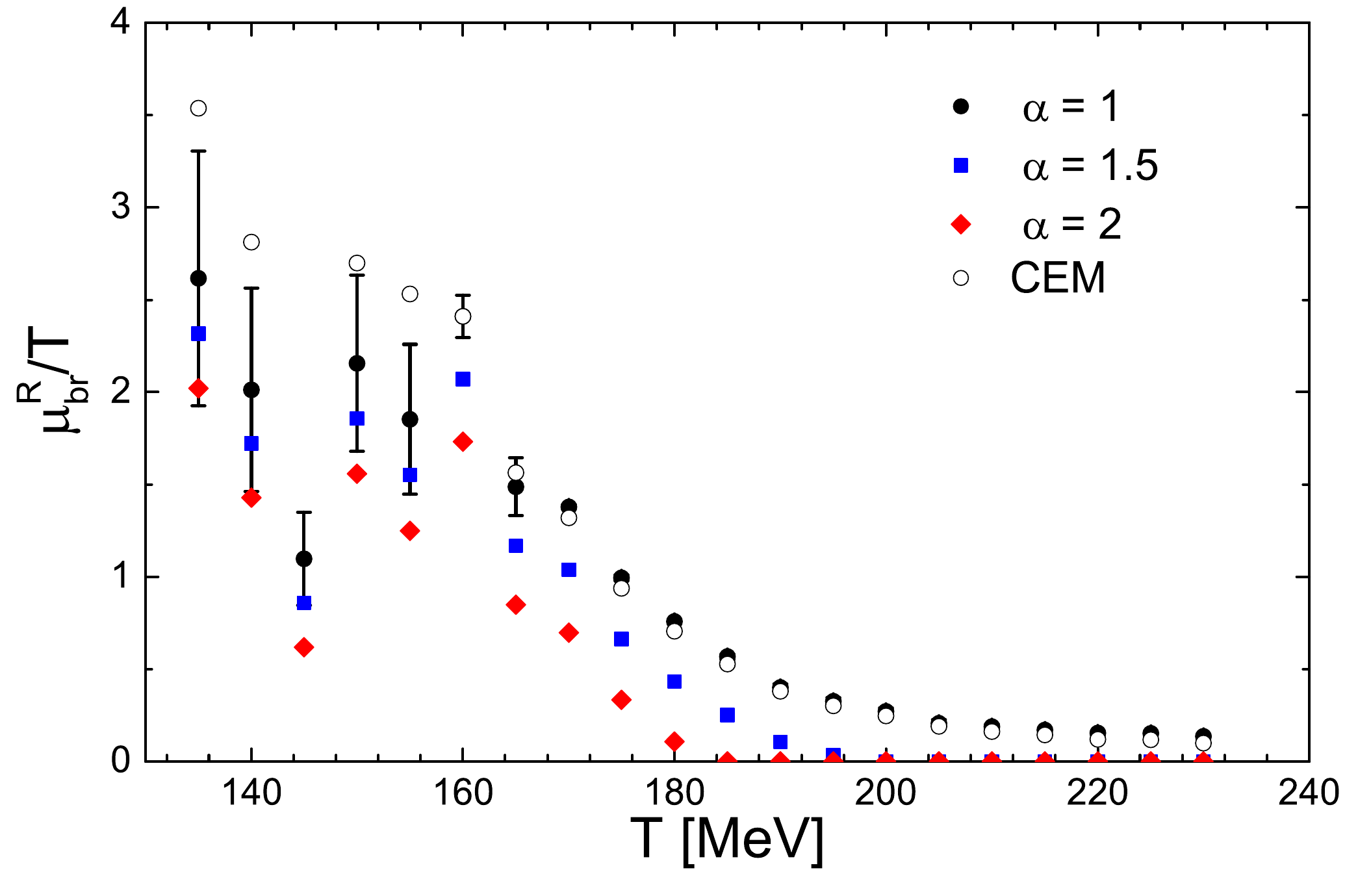}
  \caption{
   Results of the fits to the lattice data on the four leading Fourier coefficients with the ansatz~\eqref{eq:fitansatz}.
   Left panel depicts the lattice data for $\ln |b_k|$~(orange symbols), as well as fit results using $\alpha = 1$~(black lines), $\alpha = 3/2$~(blue lines), and $\alpha = 2$~(red lines), for three different temperatures~(from bottom to top): $T = 135$, 170, and 215~MeV.
   The full and open symbols corresponds to positive and negative values of $b_k$, respectively.
   The right panel shows the temperature dependence of the extracted values of $\mu_{\rm br}^{R}(T)$ for the three values of $\alpha$. Predictions for $\mu_{\rm br}^{R}$ from the cluster expansion model~\cite{Vovchenko:2017gkg} are shown by the open black symbols for comparison.
   For presentation purposes, the error bars are shown only for $\alpha = 1$, for all other cases they are of similar magnitude.
  }
  \label{fig:lqcdfits}
\end{figure*}

The exercise shows that even only four leading Fourier coefficients might be sufficient to extract the real part $\mu_{\rm br}^R$ of the limiting thermodynamic singularity under certain circumstances.
Unfortunately, the method yields no conclusive answer with regards to the nature of the extracted singularity, in particular to the possible presence of an imaginary part.
Extraction of $\mu_{\rm br}^{\rm I}$ requires an analysis of the possible oscillatory behavior of $b_k$'s which might require knowledge of a considerably larger number of coefficients.
Nevertheless, the extracted $\mu_{\rm br}^R$ in all likelihood serves as a lower bound on the value of the critical chemical potential at a given temperature.
Indeed, a presence of a singularity which is closer to the imaginary axis would imply a weaker exponential damping of Fourier coefficients.

We now repeat the fit procedure using the real lattice QCD data for $b_1$, $b_2$, $b_3$, and $b_4$~\cite{Vovchenko:2017xad} for the temperature range $135 < T < 230$~MeV. Results are depicted in Fig.~\ref{fig:lqcdfits}.
In addition to $\alpha = 3/2$, here we also consider two additional cases: $\alpha = 1$ and $\alpha = 2$.
This allows to asses the sensitivity of the results to the choice of $\alpha$, which can be different depending on the nature of singularity.
While $\alpha \approx 3/2$ is expected for a (crossover) phase transition at finite baryon density, one can also have $\alpha = 1$ in the case of a Roberge-Weiss transition~\cite{Kashiwa:2017swa} or $\alpha \approx 2$ for a chiral crossover singularity for small quark masses~\cite{Almasi:2019bvl}.
These considerations lead us to assume the interval $1 \leq \alpha \leq 2$ as a reasonable bound on $\alpha$.

The left panel of Fig.~\ref{fig:lqcdfits} depicts the fit results for three different temperatures: $T = 135$, 170, and 215~MeV.
One can see that the exponential suppression is stronger at lower temperatures. 
At $T = 135$~MeV fits are not sensitive to the chosen value of $\alpha$.
At $T = 175$~MeV the differences are seen more clearly:
while $\alpha = 1$ and $\alpha = 3/2$ fits still describe the data reasonably well, $\alpha = 2$ does a noticeably worse job.
For $T = 215$~MeV, the $\alpha = 2$ case appears to be ruled out.
The temperature dependence of the extracted $\mu_{\rm br}^{\rm R}/T$ values is depicted in the right panel of Fig.~\ref{fig:lqcdfits}.
Fits at $T \gtrsim 200$~MeV are characterized by small~($\alpha = 1$) or vanishing~($\alpha = 3/2$, 2) values of $\mu_{\rm br}^{\rm R} / T$, indicating a possibility of a power-law suppression of Fourier coefficients instead of an exponential one.
Such a scenario would correspond to a singularity at a purely imaginary value of the chemical potential. 
This may be an indication of the Roberge-Weiss transition at imaginary chemical potential at $T > T_{\rm RW}$~\cite{Roberge:1986mm}, where $T_{\rm RW} \simeq 208$~MeV according to lattice QCD estimates~\cite{Bonati:2016pwz}.

For comparison, we also depict in Fig.~\ref{fig:lqcdfits} the predictions 
of the cluster expansion model~(CEM) of Ref.~\cite{Vovchenko:2017gkg}.
The CEM describes the available lattice data on $b_k$ within errors, the asymptotic behavior of the Fourier coefficients in this model matches the general form given in Eq.~\eqref{eq:bkgen} with $\alpha = 1$.
This model predicts $\mu_{\rm br}^{\rm R} / T = \ln\left| \hat{b}_1(T) / \hat{b}_2(T)  \right|$ with $\hat{b}_{1,2}(T)$ being the lattice data for the two leading Fourier coefficients scaled by the high-temperature Stefan-Boltzmann limiting values.
The CEM predictions agree quite well with fit results performed for $\alpha = 1$, especially at larger temperatures, as seen by comparing the open and full circles in Fig.~\ref{fig:lqcdfits}.

\section{Discussion and conclusions}

We have determined properties of the cluster expansion in fugacities 
that are associated with a presence of a phase transition and a critical point at finite baryon density.
This has been achieved through the trivirial model~(TVM) -- a model which does contain a phase transition of a liquid-gas type and where one can evaluate the cluster expansion coefficients explicitly~[Eq.~\eqref{eq:bksol}].

The nontrivial behavior of $b_k$ associated with the phase transition present in the TVM is encoded in the properties of Hermite polynomials. 
The asymptotic behavior of $b_k$ changes qualitatively as one traverses the critical temperature: at $T < T_c$ one observes a monotonic behavior characterized by an exponential suppression of $b_k$ at large $k$~[Eq.~\eqref{eq:bkTmTc}], which is superimposed on a power-law damping.
At $T = T_c$ the behavior is similar, with a modification to the power-law exponent~[Eq.~\eqref{eq:bkTTc}].
The magnitude of the exponential suppression at $T = T_c$ is determined by the value of the critical chemical potential $\mu_c$ which corresponds to the critical point, $b_k \sim e^{-k \mu_c/T_c}$.
At $T > T_c$ the thermodynamic branch points move into the complex $\mu_B$ plane, which leads to an emergence of an oscillatory behavior in addition to the exponential and power-law decays in magnitude~[Eq.~\eqref{eq:bkTpTc}].
An appearance of negative values of $b_k$ above a certain temperature may therefore signal reaching the crossover temperature region, $T > T_c$.

In all cases, the asymptotic behavior of $b_k$ is determined in the TVM by the location of the phase transition branch points, and is given in most general case by Eq.~\eqref{eq:bkgen}.
Given the universality of the critical behavior, our results obtained in the framework of the TVM are expected to be qualitatively generic for any critical endpoint of a phase transition.
In Appendix B we supplement these results with similar findings obtained within a Nambu-Jona-Lasinio description in a mean-field approximation.

In QCD, the cluster expansion properties are particularly interesting in the context imaginary chemical potentials.
There, the cluster expansion coefficients become Fourier expansion coefficients of net baryon density. First-principle lattice QCD simulations are free of sign problem at imaginary $\mu_B$, 
and a calculation of a number of the leading Fourier coefficients appears to be feasible.
In fact, the four leading coefficients have already been computed on the lattice for temperatures $135 < T < 230$~MeV, although the error bars at the smaller temperatures of that range are still quite large.

According to Eq.~\eqref{eq:bkgen}, the magnitude of $b_k$ drops exponentially with $k$, with the slope proportional to the real part $\mu_{\rm br}^{\rm R}/T$ of the chemical potential of the phase transition branch point, be it a spinodal point at $T < T_c$, a critical point at $T = T_c$, or a crossover branch point at $T > T_c$.
An analysis of this exponential suppression appears to be a fairly reliable way to extract the value of $\mu_{\rm br}^{\rm R}/T$, even just four leading Fourier coefficients might be sufficient, as we show in Sec.~\ref{sec:fitting}.
Our analysis of the available lattice data suggests $\mu_{\rm br}^{\rm R}/T \leq 2-3$ at $T > 135$~MeV, with decreasing values at higher temperatures~(see Fig.~\ref{fig:lqcdfits}).
These values can serve as a reliable lower bound on the critical point location at $T > 135$~MeV.
Furthermore, given that the available lattice data contains negative Fourier coefficients $b_k < 0$ at all temperatures where the data are available~($135 < T < 230$~MeV), this disfavors the existence of the critical point at these temperatures.
At $T \gtrsim 200$~MeV the $\mu_{\rm br}^{\rm R}/T$ values are small, in some cases even vanishing.
This might serve as an indication of the Roberge-Weiss transition at purely imaginary $\mu_B$.

In summary, we have presented an analysis of a hypothetical phase transition and a critical point at finite baryon density using the coefficients of the cluster expansion in fugacities.
This provides a complementary approach in the hunt for the QCD critical point, in addition to the commonly used methods based on conserved charges susceptibilities, which are either calculated at $\mu_B = 0$ in lattice QCD or measured at non-zero $\mu_B$ in heavy-ion collisions.
The results obtained are useful for future lattice QCD simulations at imaginary $\mu_B$, which will hopefully yield more accurate values of Fourier coefficients at temperatures where the QCD critical point can be expected.
Analysis of the structure of these Fourier coefficients will be able to yield new bounds on the possible location~(or even existence) of the QCD critical point.

%TC:ignore 

\begin{acknowledgments}

%\section*{Acknowledgements}
%\emph{Acknowledgments.} 
We thank Szabolcs Bors\'anyi and Bengt Friman for fruitful discussions and useful comments.
H.St. acknowledges the support through the Judah M. Eisenberg Laureatus Chair by Goethe University  and the Walter Greiner Gesellschaft, Frankfurt. V.K. is supported by the U.S. Department of Energy, Office of Science, Office of Nuclear Physics, 
under contract number DE-AC02-05CH11231. 
This work was supported by the DAAD through a PPP exchange grant.
This work also received support within the framework of the Beam Energy Scan Theory (BEST) Topical Collaboration.

\end{acknowledgments}

\section*{Appendix}

\subsection{Trivirial model as an equation of state of a real gas}
\label{app:A}

The TVM introduced in this paper~(Sec.~\ref{sec:tvm}) can be viewed as a variant of a real gas equation of state,
constructed for a system of particles with short-range repulsive~(excluded volume) and intermediate range attractive~(mean field) interactions.
A generic framework of real gas models, including the effects of quantum statistics, was developed in Ref.~\cite{Vovchenko:2017cbu} and applied to model the nuclear matter.
The free energy of a real gas in this framework is the following:
\eq{\label{eq:Frg}
F(T,V,N) = F^{\rm id}(T, V \, f(\eta), N) + N \, u(n).
}
Here $f(\eta)$ is an available volume fraction. It models the short-range repulsive interactions in a form of a generalized excluded volume procedure. 
$\eta \equiv (bN) / 4V$ and $b$ is the excluded volume parameter.
$u(n)$ with $n \equiv N / V$ is an attractive mean-field.
Comparing Eq.~\eqref{eq:Frg} with the free energy expression \eqref{eq:F} in the TVM allows to obtain the explicit TVM expressions for $f(\eta)$ and $u(n)$:
\eq{
f^{\rm tvm} (\eta) & = \exp\left( -4 \eta - 8 \eta^2 \right), \\
u^{\rm tvm} (n)    & = - a \, n.
}

The real gas formulation of the TVM brings new possible applications.
For example, the TVM can be used to model nucleon-nucleon interactions and the nuclear liquid-gas transition.
Following the generic procedure described in Ref.~\cite{Vovchenko:2017cbu} one can fix the parameters $a$ and $b$ of nucleons to reproduce the saturation density $n_0 = 0.16$~fm$^{-3}$ and the binding energy per nucleon $E/A = -16$~MeV, yielding 
\eq{\label{eq:ab}
a^{\rm tvm} \simeq 349~\text{MeV fm}^3, \qquad b^{\rm tvm} \simeq 4.28~\text{fm}^3~
}
for the TVM. The model predicts a critical point of nuclear matter at
\eq{
& T_c^{\rm tvm} \simeq 18.3~\text{MeV}, \qquad n_c^{\rm tvm} \simeq 0.07~\text{fm}^{-3}, & \nonumber \\ 
& \mu_c^{\rm tvm} \simeq 910~\text{MeV}~, &
}
which is in a reasonable agreement with empirical estimates~\cite{Elliott:2013pna}.
The TVM can also be used to incorporate the baryon-baryon interactions and the associated nuclear liquid-gas criticality into a hadron resonance gas model, in a similar way as it was done in Ref.~\cite{Vovchenko:2016rkn} for the vdW equation.

\subsection{Fourier coefficients in an NJL model}
\label{app:NJL}

This Appendix presents the behavior of Fourier coefficients in an NJL model~\cite{Nambu:1961tp,Nambu:1961fr}.
The NJL model is a low-energy effective theory of QCD~\cite{Buballa:2003qv}, which has been used in the past to study the phase structure of QCD, in particular that associated with the critical behavior~\cite{Hatsuda:1994pi,Asakawa:2009aj}. 
The model exhibits a chiral critical point and a first-order phase transition at finite net quark number densities.
Therefore, it can be interesting to consider the behavior of the Fourier coefficients in NJL, in particular to verify the asymptotic behavior~\eqref{eq:bkTmTc}-\eqref{eq:bkTpTc} of $b_k$ associated with the critical point, which was obtained in the framework of the TVM and which we expect to be model-independent.

We take a mean-field variant of the NJL model for 2 flavors and 3 colors and neglect the vector repulsion.
The quark chemical potential $\mu_q$ plays the role of the chemical potential $\mu$.
The grand potential reads~\cite{Buballa:2003qv}
\eq{
& \Omega(T,\mu;M) = -\frac{12}{2\pi^2} \, \int_{0}^{\Lambda} \, k^2 \, dk \, \left\{
\sqrt{k^2+M^2} \right. \nonumber \\
& \quad + T \ln\left[1 + \exp\left(-\frac{\sqrt{k^2+M^2}-\mu}{T}\right) \right] 
\nonumber \\
& \quad \left. + T \ln\left[ 1 + \exp\left(-\frac{\sqrt{k^2+M^2}+\mu}{T}\right) \right] \right\} \,  \nonumber  \\
& \quad + \frac{(M-m_0)^2}{4G_S}.
}

The model parameters are the momentum cut-off $\Lambda$, the bare quark mass $m_0$, and the scalar coupling $G_S$.
The constituent quark mass $M$ at given $T$ and $\mu$ is determined by minimizing the grand potential. This is defined by the gap equation:
\eq{\label{eq:NJLgap}
\frac{\partial \Omega}{\partial M} = 0.
}

\begin{figure}[t]
  \centering
  \includegraphics[width=.49\textwidth]{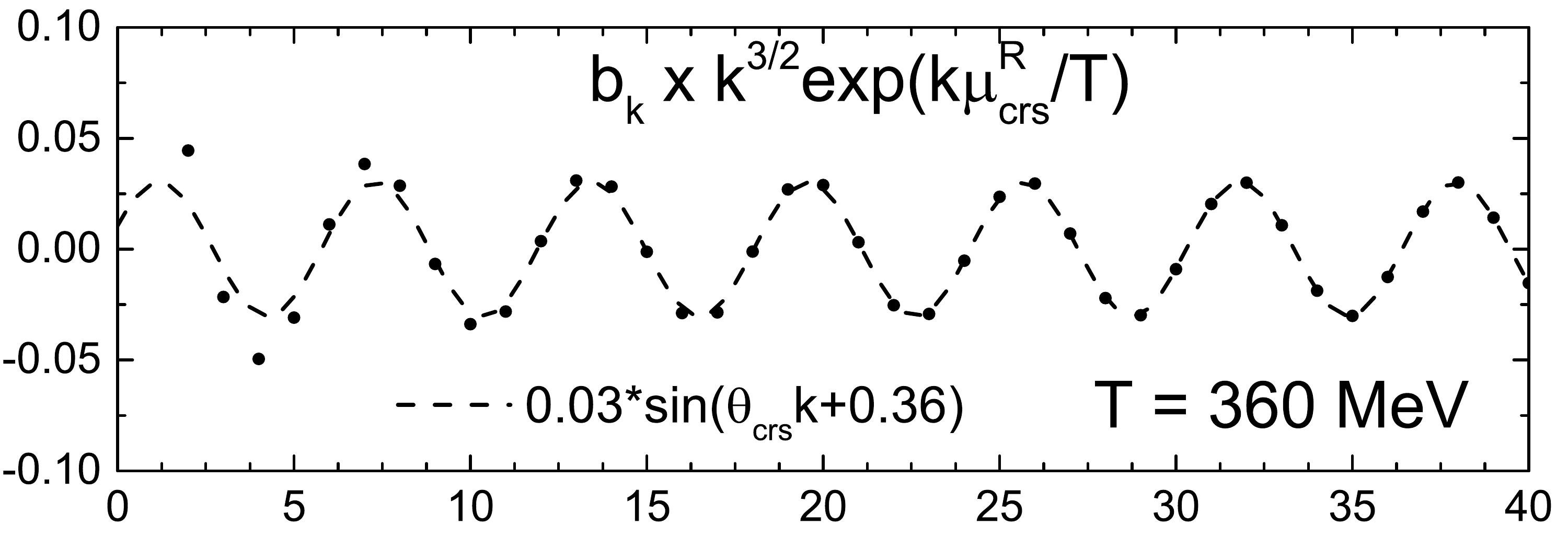}
  \includegraphics[width=.49\textwidth]{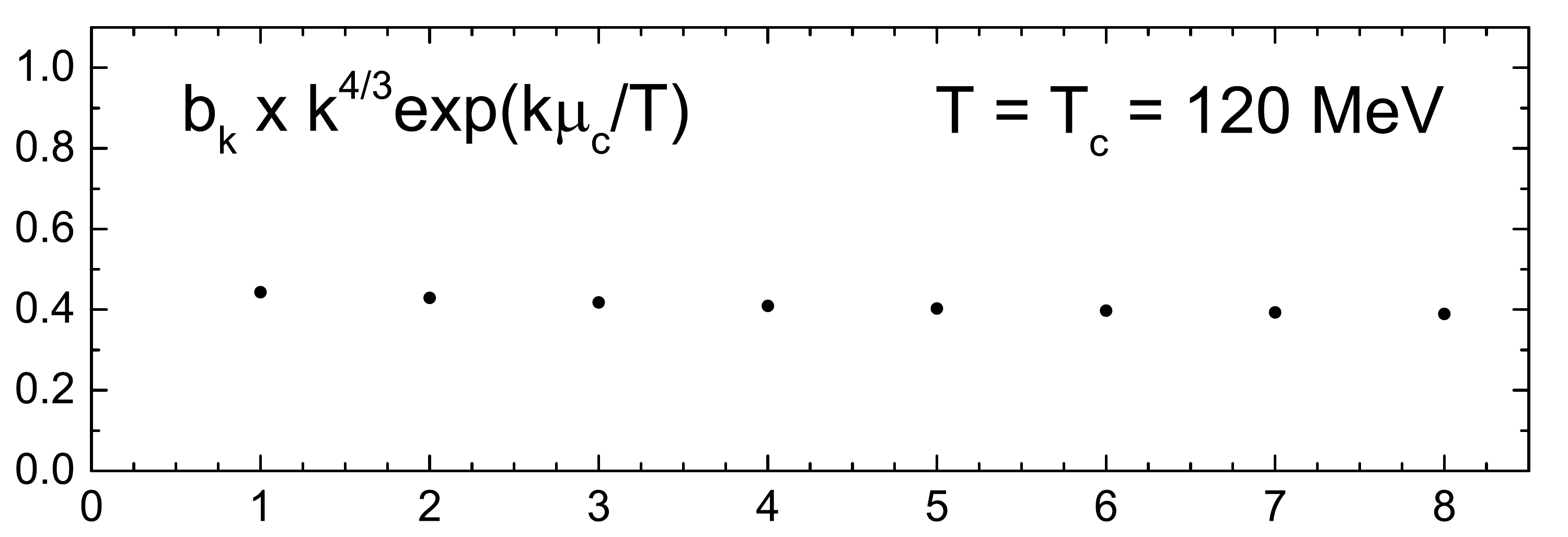}
  \includegraphics[width=.49\textwidth]{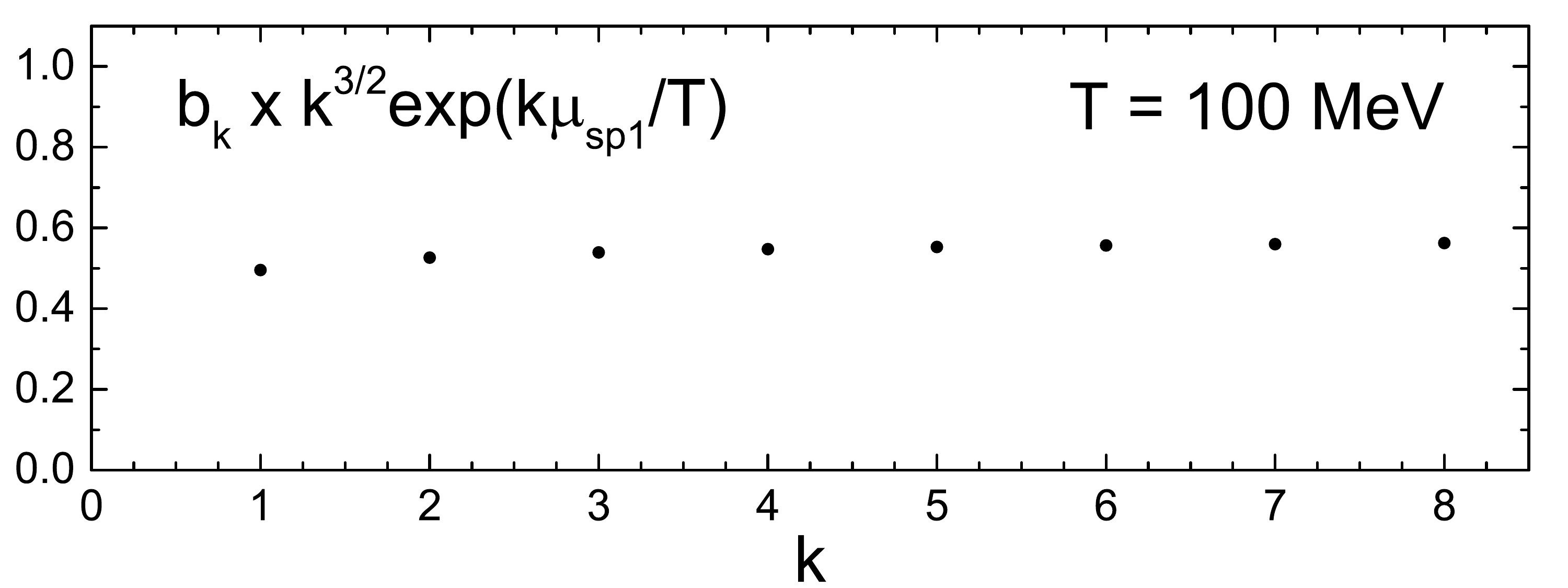}
  \caption{
  Behavior of the Fourier coefficients $b_k$
  in an NJL model for three different temperatures~(from top to bottom): a supercritical temperature of $T = 360$~MeV, the critical temperature, $T = T_c = 120$~MeV, and a subcritical temperature of $T = 100$~MeV.
  The coefficients are scaled by the expected power-law and exponential factors.
  The dashed line in the top panel shows the fit of the scaled $b_k$ coefficients with an oscillatory function from~Eq.~\eqref{eq:bkTpTc} with $\theta_{\rm crs} \equiv \mu_{\rm crs}^I / T \simeq 1.03$.
  }
  \label{fig:NJLbk}
\end{figure}

Two solutions of the gap equation may merge at a branch point. The branch points are defined from
\eq{\label{eq:NJLbr}
\frac{\partial^2 \Omega}{\partial M^2} = 0,
}
and have the same physical meaning as the thermodynamic branch points introduced in Sec.~\ref{sec:tvm} within the TVM.
The branch point coordinates at a given temperature can be determined in the NJL model by solving numerically Eqs.~\eqref{eq:NJLgap} and \eqref{eq:NJLbr}.

Two branch points merge at the critical point. This corresponds to
\eq{\label{eq:NJLCP}
\frac{\partial^3 \Omega}{\partial M^3} = 0.
}
Equations~\eqref{eq:NJLgap}-\eqref{eq:NJLCP} determine the location of the critical point in the NJL model.
Here we take the parameter set 3 from Ref.~\cite{Buballa:2003qv}, namely $\Lambda = 569.3$~MeV, $G_S = 2.81 / \Lambda^2$, and $m_0 = 5.5$~MeV.
The critical point location is
\eq{
T_c = 120~\text{MeV}, \qquad \mu_c = 348~\text{MeV}.
}

The behavior of Fourier coefficients $b_k$ is studied by computing the Fourier integrals of net quark number density at imaginary $\mu$ through a numerical integration:
\eq{
b_k(T) = \frac{2}{\pi} \int_0^{\pi} \text{Im} \left[ \frac{\rho(T, i \theta \, T)}{T^3} \right] \, \sin(k \, \theta) \, d \theta~.
}
The evaluation of the net quark number density $\rho(T,\mu)$ at imaginary $\mu$ is done in two steps.
First, the effective mass $M$ at a given $\mu$ is computed by minimizing the grand potential $\Omega$. This is achieved by solving the gap equation~\eqref{eq:NJLgap}.
Then, the density is computed as $\rho(T,\mu) = -(\partial{\Omega}/\partial{\mu})_T$.

Calculation results for $b_k$ are depicted in Fig.~\ref{fig:NJLbk}, for three different temperatures: a supercritical temperature of $T = 360$~MeV, the critical temperature, $T = T_c = 120$~MeV, and a subcritical temperature of $T = 100$~MeV.
The coefficients are scaled by the expected asymptotic power-law and exponential factors from Eqs.~\eqref{eq:bkTmTc}-\eqref{eq:bkTpTc}, for which we determine $\mu_{\rm crs}^{R}$, $\mu_c$, and $\mu_{\rm sp1}$ numerically, by solving Eqs.~\eqref{eq:NJLgap} and~\eqref{eq:NJLbr}:
$\mu_{\rm crs}^{R} \simeq 47.3$~MeV for $T = 360$~MeV, $\mu_c \simeq 348$~MeV for $T = T_c = 120$~MeV, and $\mu_{\rm sp1} \simeq 368$~MeV for $T = 100$~MeV.
The scaled coefficients quickly flatten for $T < T_c$ and $T = T_c$~(the two lower panels in Fig.~\ref{fig:NJLbk}) whereas at $T>T_c$ they show an oscillatory behavior~(dashed line in Fig.~\ref{fig:NJLbk}), as predicted by the TVM~[Eq.~\eqref{eq:bkTpTc}].
The numerical NJL model results thus confirm the analytic TVM predictions for the asymptotic behavior of $b_k$.

\bibliography{tvm}

%TC:endignore 

\end{document}